%% file: main.tex
\documentclass{article}
\usepackage{spconf,amsmath,graphicx}
\usepackage{hyperref}
\usepackage{multirow}
\usepackage{stfloats}
\usepackage{booktabs}
\usepackage{bm}
\usepackage{enumitem}
\usepackage{threeparttable}
\usepackage{pifont}
\usepackage{xcolor}

\title{Consistent and Relevant: Rethink the query embedding in \\general sound separation}

\name{
\begin{tabular}{c}
Yuanyuan Wang$^{1,\ast}$, Hangting Chen$^{2}$, Dongchao Yang$^3$, Jianwei Yu$^{2}$, Chao Weng$^{2}$,\\ Zhiyong Wu$^{1,3,\dagger}$, Helen Meng$^3$
\end{tabular}
\thanks{$^{\ast}$Work performed during an internship at Tencent AI Lab.}
\thanks{$^\dagger$Corresponding author.}
}
\address{
 $^1$ Shenzhen International Graduate School, Tsinghua University, Shenzhen, China\\
  $^2$ Tencent AI Lab, Audio and Speech Signal Processing Oteam, China\\
  $^3$ The Chinese University of Hong Kong, Hong Kong SAR, China\\
 \small{wangyuan21@mails.tsinghua.edu.cn, erichtchen@tencent.com, zywu@sz.tsinghua.edu.cn}
}

\begin{document}
\ninept
\maketitle
\begin{abstract}
The query-based audio separation usually employs specific queries to extract target sources from a mixture of audio signals.
Currently, most query-based separation models need additional networks to obtain query embedding. 
In this way, separation model is optimized to be adapted to the distribution of query embedding.
However, query embedding may exhibit mismatches with separation models due to inconsistent structures and independent information.
In this paper, we present CaRE-SEP, a consistent and relevant embedding network for general sound separation to encourage a comprehensive reconsideration of query usage in audio separation. 
CaRE-SEP alleviates the potential mismatch between queries and separation in two aspects, including sharing network structure and sharing feature information.
First, a Swin-Unet model with a shared encoder is conducted to unify query encoding and sound separation into one model, eliminating the network architecture difference and generating consistent distribution of query and separation features.
Second, by initializing CaRE-SEP with a pretrained classification network and allowing gradient backpropagation, 
the query embedding is optimized to be relevant to the separation feature, 
further alleviating the feature mismatch problem. 
Experimental results indicate the proposed CaRE-SEP model substantially improves the performance of separation tasks. Moreover, visualizations validate the potential mismatch and how CaRE-SEP solves it.

\end{abstract}

\input{01_introduction}
\input{02_method}

\input{03_experiments}
\input{04_conclusions}

\vfill\pagebreak

\bibliographystyle{IEEEbib}
\bibliography{strings,refs}

\end{document}

%% file: 01_introduction.tex
\vspace{-5pt}
\section{introduction}
\label{sec:introduction}
\vspace{-3pt}





Audio source separation is a process of separating one or more isolated sound sources from a complex auditory scene \cite{nugraha2016multichannel, 8683007}.
It plays a crucial role in many downstream tasks such as audio extraction, audio transcription, music editing, and speech enhancement \cite{vincent2006performance, nugraha2016multichannel, chen2022zero, wang2022improving}. 
In recent years, the rapid progress in deep learning technologies has largely improved audio source separation performance~\cite{jansson2017singing,stoller2018wave, 8462116}. 
Many methods~\cite{7760548,7952158,8521383} train separate models for different target types of audio sources, which are effective for source separation.
However these models require significant computational resources and training data, making it challenging to expand the number of possible audio sources, 
and so they can only separate the fixed number of sources~\cite{8683007, 8683800}.

To tackle the problem of constrained source types, some query-based separation models~\cite{lee2019audio, chen2022zero} have been developed. 
Lee \textit{et al.}~\cite{lee2019audio} encoded the specific target source embedding with an extra Query-net, to separate out a target source from an audio mixture.
Recently, Chen \textit{et al.}~\cite{chen2022zero} designed a Transformer-based audio classification model and a CNN-based separation model.
They then use these models to extract a target source from an audio mixture given a target source embedding that is obtained from the pre-trained audio classification model, which is trained for audio tagging tasks.

The above query-based audio separation models aim to separate as many sources as possible, but there may be some structural and informational mismatches between query embedding and separation feature. 
Firstly, these methods require query embedding from an auxiliary model as query.
Although these auxiliary models can distinguish different patterns and features in the data to some extent,
the auxiliary structure is inconsistent with the encoder of separation network, which may cause a mismatch between the distribution of query embedding and separation feature.
For example, Transformer pays attention to global dependencies and CNN focuses more on local features. 
Besides, it is important to note that the auxiliary network is directly optimized for other tasks, such as audio classification. So the query embedding obtained from it may not be directly relevant to the primary separation task.
Thus, query embedding and separation feature might encounter information mismatch. 

In this paper, we propose a novel Consistent and Relevant Embedding network for audio source separation (CaRE-SEP) by eliminating structural and informational mismatches from the point that the impact of query embedding in query-based audio separation. 
Firstly, inspired by \cite{cao2022swin}, we propose a Swin-Unet model incorporating a shared encoder to integrate query encoding and sound separation within a single model.
Owing to the shared encoder module, the distribution of generated query embedding and separation feature are also more consistent, which can contribute to the separation task. 
Subsequently, we initialize the encoder of CaRE-SEP with a pre-trained classification network and dynamically train the query embedding with gradients.
This contributes to query embedding being more relevant to the separation task, further boosting the separation performance.
Besides, the shared encoder can achieve audio classification, so our CaRE-SEP can jointly train classification and separation tasks.
In summary, the main contributions of our CaRE-SEP are as follows:

\begin{itemize}[itemsep=0pt,topsep=0pt,parsep=0pt,leftmargin=10pt]
\item 
\textbf{Consistent distribution}: 
Since query embedding and separation feature are generated by a shared encoder, their distributions are consistent, which is advantageous for audio separation tasks.
\item 
\textbf{Relevant information}: The query embedding is more relevant to audio source separation with gradient and initialization, which also simplifies the model structure.
\item 
\textbf{Unified structure}: We are the first to propose a unified model that can jointly train classification and separation tasks, which indicates that a single architecture can be simultaneously utilized for classification and separation tasks.
\end{itemize}

%% file: 02_method.tex
\vspace{-7pt}
\section{Method}
\label{sec:method}
\begin{figure}[htb]
\centering
\includegraphics[width=8.6cm]{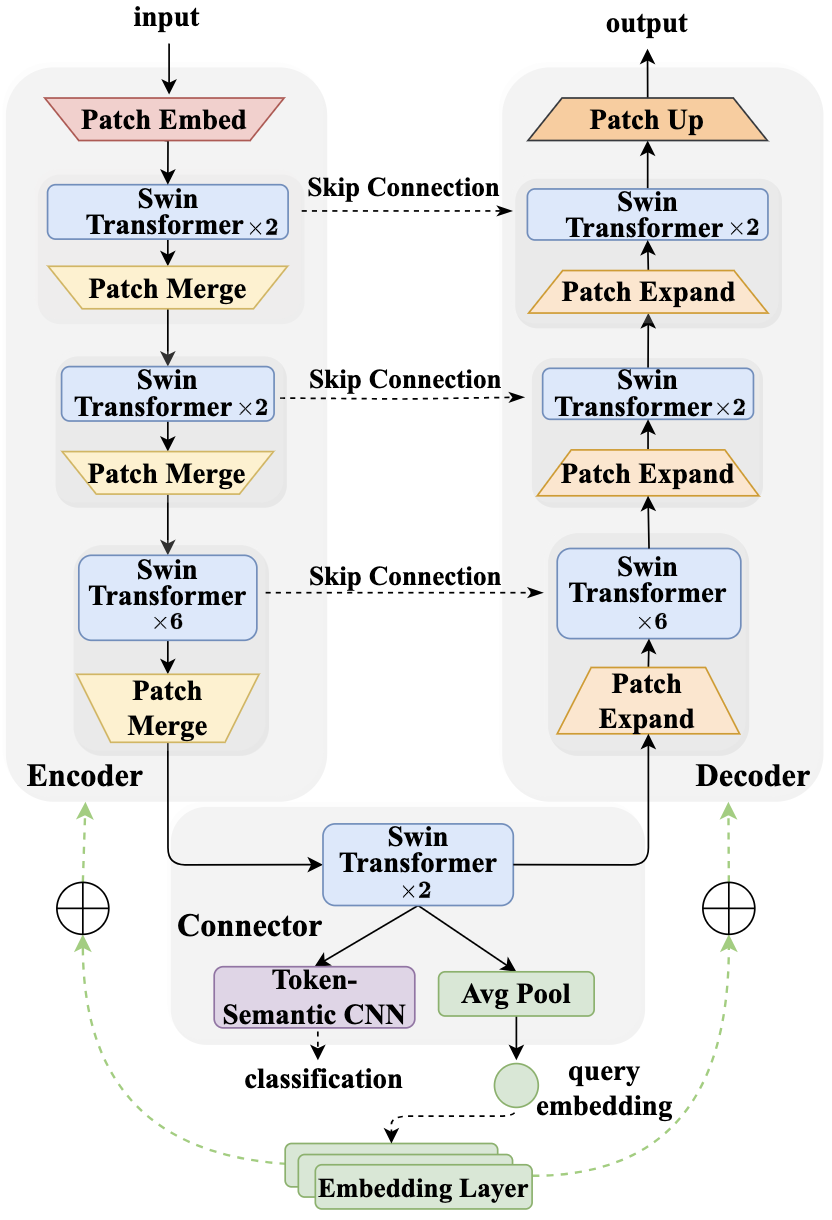}
\vspace{-20pt}
\caption{Consistent and Relevant Embedding Separation Networks (CaRE-SEP)}
\label{fig:detail}
\vspace{-20pt}
\end{figure}

In this section, we present the Consistent and Relevant Embedding network for audio source separation (CaRE-SEP).
Fig.\ref{fig:overall} and Fig.\ref{fig:detail} illustrate the overall and detailed network, respectively. 

\vspace{-10pt}
\subsection{Overall architecture}
\begin{figure}[htb]
\vspace{-8pt}
\centering
\includegraphics[width=8.3cm]{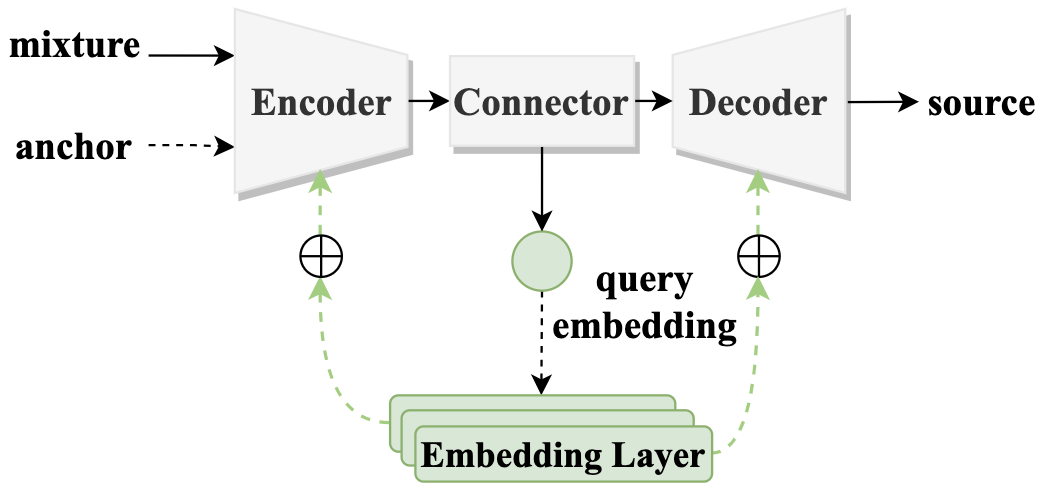}
\vspace{-10pt}
\caption{The overall architecture}
\label{fig:overall}
\vspace{-6pt}
\end{figure}

As shown in Fig.\ref{fig:overall}, our CaRE-SEP is composed of shared encoder, connector module and decoder module, where the encoder and connector module are responsible for query embedding, and all of them are trained for the audio source separation.

In Fig.\ref{fig:overall}, we regard query embedding network as a part of separation model and use a shared encoder module to simultaneously obtain query embedding and separation feature, which is different from \cite{chen2022zero} that uses a frozen extra classification model (ST-SED) to obtain query embedding. 
When the query embedding module and separation structure are similar, the distribution of generated query embedding and separation feature will be more consistent, leading to optimal performance. 
Furthermore, we initialize the shared encoder of CaRE-SEP with a pre-trained classification model~\cite{chen2022hts} and perform gradient propagation for query embedding. 
As a result, the query embedding is optimized to be more relevant to separation task. 
Next, we will introduce the training process depicted in Fig.\ref{fig:overall}.


We define two audio sources $a_1, a_2$ and mix them $a = a_1 + a_2$ with energy normalization. 
To balance classification and separation performance, we do not use mel-spectrograms as input which perform worse in separation.
Instead, we use Mel scale-based band-split \cite{chen23t_interspeech} to split the spectrogram into a series of subband spectrograms.
The mixture audio $a$ is converted into the
spectrogram by Short-time Fourier Transform (STFT) and then split into subband spectrograms. 
Besides, we use anchor audios to obtain fixed-dimensional query embeddings $e_1, e_2$, which are incorporated into the CaRE-SEP to specify which audio needs to be separated. This process is illustrated by the dashed line in Fig.\ref{fig:overall}.
Subsequently, we send two training triplets $(a, a_1, e_1),(a, a_2, e_2)$ into the CaRE-SEP model S, respectively. Our separation objective is as follows:
\begin{equation}\label{target}
\vspace{-3pt}
\begin{aligned}
S(a, e_j) \rightarrow a_j, j \in {1, 2}
\end{aligned}
\vspace{-2pt}
\end{equation}
Simultaneously, the linear layer (embedding layer in Fig.\ref{fig:overall}) is applied on the query embedding $e_j$ to unify the feature dimension, and then the queries are added into the audio feature maps of each layer block in the CaRE-SEP model, helping to specify the audio that needs to be separated.
As a result, the model will also learn the relationship between the query embedding and mixture, adjusting its weights to adapt to the separation feature. 
Finally, the output spectrogram is converted into the waveform by inverse STFT (iSTFT).
We apply the Mean Absolute Error (MAE) and Source-to-Distortion Ratio (SDR) to compute the loss between separate waveforms and the target source.


\vspace{-8pt}
\subsection{Shared encoder module}
As illustrated in the left part of Fig.\ref{fig:detail}, our shared encoder mainly
consists of Swin Transformer \cite{liu2021swin} and Patch Merge  \cite{chen2022hts, chen2022zero}. 
The Patch-Embed CNN splits the input features into different patch tokens inside each window, which more effectively capture the relationship among frequency bins of the same time
frame.
The transformed patch tokens are sent into three groups of Swin Transformer blocks \cite{liu2021swin} and Patch-Merge layers to generate hierarchical representations.
Within each group, we adopt the more efficient swin transformer block with a shifted window attention to learn feature representation.
Specifically, the patch-merge layer is applied to decrease the sequence size at the end of each group. 

\vspace{-5pt}
\subsection{Connector module}
The connector module contains a swin transformer, token-semantic module \cite{chen2022zero}, and average pooling layer.
Hierarchical representations of the swin transformer have the ability to address multiple tasks. 
Firstly, hierarchical representations can be input into the decoder network to separate isolated audio sources from mixed audio.
Besides, they can be processed through the average pooling layer to generate query embedding for separation. 
Additionally, the token-semantic CNN integrates all frequency bins and maps the feature into the class probabilities.
The vectors produced by the token-semantic CNN can accomplish audio tagging tasks by computing binary cross-entropy loss with ground truth labels.
So we can simultaneously train tagging and separation tasks in a single model.
\vspace{-10pt}
\subsection{Decoder module}
Taking inspiration from U-Net \cite{ronneberger2015u}, the decoder of our CaRE-SEP is a structure symmetrical to our encoder, which can better accomplish separation tasks.
It is composed of three Swin Transformer blocks and three patch expand layers.
Different from the patch merge layer, the patch expand layer is devised for up-sampling, which transforms adjacent feature maps into larger feature maps with 2x up-sampling of resolution.
Finally, the patch up layer performs 4x up-sampling to restore the feature maps to the original input resolution.
The skip connection combines extracted representations of the decoder and multi-scale features from the encoders together, which helps to complement the loss of spatial information resulting from down-sampling.


%% file: 03_experiments.tex
\vspace{-5pt}
\section{experimental setup}
\vspace{-8pt}
\subsection{Dataset}
\textbf{AudioSet.} Our CaRE-SEP model is trained on AudioSet \cite{gemmeke2017audio}, a large-scale dataset of over two million 10-second audio samples labeled with 527 sound event categories. 
In the following experiments, we use the full-train set of AudioSet (2M samples) to train our model and evaluate its performance on the evaluation set (22K samples).

\noindent \textbf{ESC-50} \cite{piczak2015esc}.
To further evaluate the generalization ability of the separation task, we simulate a mixed audio test dataset based on ESC-50 dataset.
The ESC-50 
contains 5-second-long recordings distributed across 50 semantic classes.
For each audio in a specific category, we randomly select an audio from another category and mix them.
As a result, we generate 2,000 simulated mixture samples.
\vspace{-10pt}
\subsection{System Implementation}
\noindent \textbf{Implementation details.} 
We use 1024 window size and 320 hop size to compute spectrograms. We use Mel scale-based band-split \cite{chen23t_interspeech} to compress the spectrogram into 64 bands.
For Swin Transformer block, we share the same configuration with \cite{chen2022hts}.
There are two options for latent dimension size D=96 or D=256 corresponding to query embedding of 8D=768 or 8D=2048.
Our CaRE-SEP is implemented in Pytorch and trained with AdamW optimizer ($\beta_1$=0.9, $\beta_2$=0.999, eps=1e-8, decay=5e-4) and a batch size of 96.
We utilize a warm-up schedule, initially setting the learning rate at 0.05, 0.1, and 0.2 for the first three epochs. Subsequently, the learning rate is reduced by half every ten epochs until it reaches 0.05.

\noindent \textbf{System evaluation.} 
For the audio tagging task, we use mean average precision (mAP) as the metric. 
We use Source-to-Distortion Ratio (SDR) as the evaluation metric for audio separation.  
Similar to \cite{chen2022zero}, we have three SDR metrics for the validation set:
\begin{itemize}[itemsep=0pt,topsep=0pt,parsep=0pt,leftmargin=10pt]
\item 
mixture-SDR: 
$S(c_1 + c_2, e_j) \rightarrow c_j$
\item 
clean-SDR: 
$S(c_j, e_j) \rightarrow c_j$
\item 
silence-SDR: 
$S(c_{\neg j}, e_j) \rightarrow \textbf{0}$
\end{itemize}
where the $\neg j$ denotes any clip that does not have the same class as the $j$-th clip. 
The clean SDR aims to confirm the model's ability to preserve the clean source when provided with the self-source query embedding. 
The silence SDR is to validate if the model can separate nothing when the given audio lacks a target source.
\vspace{-5pt}
\section{Results and Analyses}
\vspace{-5pt}
\subsection{Results on AudioSet}\label{sec_41}
\vspace{-20pt}
\begin{table}[htbp]
\centering
\setlength{\tabcolsep}{1.8mm}
\caption{Results on AudioSet Evaluation Set. CLS refers to audio classification and SEP is audio separation. NP means not provided.}
\begin{tabular}{ccccc}
\toprule[1pt]
\multirow{2}{*}{\textbf{Model}} 
&
\multirow{2}{*}{\textbf{mAP}$\uparrow$}
& \multicolumn{3}{c}{\textbf{SDR}(dB$\uparrow$)}
\\
&
      & \multicolumn{1}{c}{Mixture} & \multicolumn{1}{c}{Clean} & \multicolumn{1}{c}{Silence} \\
\hline
AudioSet Baseline \cite{gemmeke2017audio}
& 0.314
& \text{-}
& \text{-}
& \text{-}
\\
DeepRes \cite{ford2019deep}
& 0.392
& \text{-}
& \text{-}
& \text{-}
\\
PANN \cite{kong2020panns}
& 0.434
& \text{-}
& \text{-}
& \text{-}
\\
PSLA \cite{gong_psla}
& 0.444
& \text{-}
& \text{-}
& \text{-}
\\
AST(single) \cite{gong21b_interspeech}
& 0.459
& \text{-}
& \text{-}
& \text{-}
\\
768-d ST-SED\cite{chen2022zero}
& \textbf{0.467}
& \text{-}
& \text{-}
& \text{-}
\\
\hline
527-d PANN-SEP \cite{9053396}  
 &NP
& 7.38           & 8.89         & 11.00
\\
2048-d PANN-SEP \cite{chen2022zero} 
&
NP
& 9.42           & 13.96         & 15.89       
\\
2048-d ST-SED-SEP \cite{chen2022zero}    
&
0.459
& 10.55           & 27.83         & 16.64
\\
\hline
\textbf{768-d CaRE-CLS}
& \underline{0.463}
& \text{-}
& \text{-}
& \text{-}
\\
\textbf{768-d CaRE-CLS+SEP}
& 0.446 
& 11.48
& 97.18
& \textbf{19.81}
\\
\textbf{768-d CaRE-SEP} 
&
\text{-}
& \underline{13.14}                                    &  \textbf{103.16}                                 & \underline{17.72}
\\
\textbf{2048-d CaRE-SEP}     
&
\text{-}
& 
\textbf{13.84}                                    &  \underline{96.1}                                 & 17.38
\\
\bottomrule[1pt]
\end{tabular}
\label{res_on_audioset}
\vspace{-10pt}
\end{table}

\begin{table*}[ht]
\centering
\caption{The impact of query embedding on the separation model}
\begin{tabular}{cccccccc}
\toprule[1pt]
\multirow{2}{*}{\textbf{ID}} &
\multirow{2}{*}{\textbf{Model}} &
\multicolumn{3}{c}{\textbf{Configuration}} 
& \multicolumn{3}{c}{\textbf{AudioSet Evaluation Set}}  \\
&
&
\multicolumn{1}{c}{$Grad$} 
& \multicolumn{1}{c}{$Init$}
& \multicolumn{1}{c}{$Shared\ encoder$}
      & \multicolumn{1}{c}{Mixture-SDR(dB$\uparrow$)} & \multicolumn{1}{c}{Clean-SDR(dB$\uparrow$)} & \multicolumn{1}{c}{Silence-SDR(dB$\uparrow$)} \\
\hline
A &
2048-d ST-SED-SEP \cite{chen2022zero}
&
\ding{55}
& \ding{51}
& \ding{55}
&  10.55                                   &   27.83                                &  16.64
\\
B &
w/ $Grad$
&
\ding{51}
& \ding{51}
& \ding{55}
&  10.77                                   &  31.85                                 & 13.67
\\
\hline
C &
\textbf{768-d CaRE-SEP}  & 
\ding{51}
& \ding{51}
& \ding{51}
& \textbf{13.14}                                    &  \textbf{103.16}                                 & \textbf{17.72}
\\
D &
w/o $Grad$
& 
\ding{55}
& \ding{51}
& \ding{51}
& 10.09                                     & 78.13                                   & 5.85
\\
E &
w/o $Init$
&
\ding{51}
& \ding{55}
& \ding{51}
& 11.33                                    & 100.89                                  & 15.73                                    \\
F &
w/o $Grad\&Init$
& 
\ding{55}
& \ding{55}
& \ding{51}
& 9.14            & 69.25         & 11.13           
\\
G &
w/o $Grad\&Shared$
&
\ding{55}
& \ding{51}
& \ding{55}
& 10.34                                    &  56.7                                 & 17.59
\\
\bottomrule[1pt]
\end{tabular}
\begin{tablenotes}
\footnotesize
\item $^*$Grad refers to gradient. 
Init means to initialize the encoder for query embedding. 
The shared encoder is to unify query encoding and encoder of separation.\\
$^*$w/ is with and w/o is without.
\end{tablenotes}
\label{res_ablation}
\vspace{-17pt}
\end{table*}

In this experiment, we train our CaRE-SEP model on AudioSet and evaluate the performance, with the results presented in Table \ref{res_on_audioset}. 
The hidden dimensions of the separation model have different sizes, including 527-d, 768-d, and 2048-d.

For the audio classification task, when we only train audio tagging with CaRE-CLS, the performance is almost close to the best result. 
The difference between them is that 768d ST-SED uses mel-spectrogram while we use subband spectrogram as input. 
When we perform joint training in CaRE-CLS+CEP, the classification performance is slightly worse, the separation performance is better than the AudioSet baseline but not as good as our CaRE-SEP, indicating that there are some conflicts between the classification feature and the separation feature.
We will further explain this in Section \ref{sec_distribution}

For the separation task, 
we could clearly figure out that our CaRE-SEP models trained at 768-d and 2048-d embeddings achieve optimal performance. 
Compared to the baseline (2048-d ST-SED-SEP), our best mixture-SDR increases by more than 3 dB, and the improvement of clean-SDR is quite significant, demonstrating that our model can maintain the audio features very well.
All of these results demonstrate the effectiveness of our proposed model.

It is worth noting that our model exhibits a similar performance in both the 768-d and 2048-d dimensions. 
A possible reason is that the larger embedding space may have some redundancy, and the 768-d query embedding space can already help the model better capture the audio features and provide more discriminative embeddings for more accurate separation. 

\vspace{-10pt}
\subsection{The distribution of embedding}\label{sec_distribution}

\begin{figure}[htb]
\centering
\begin{minipage}[b]{.48\linewidth}
  \centering
  \centerline{\includegraphics[width=4.0cm]{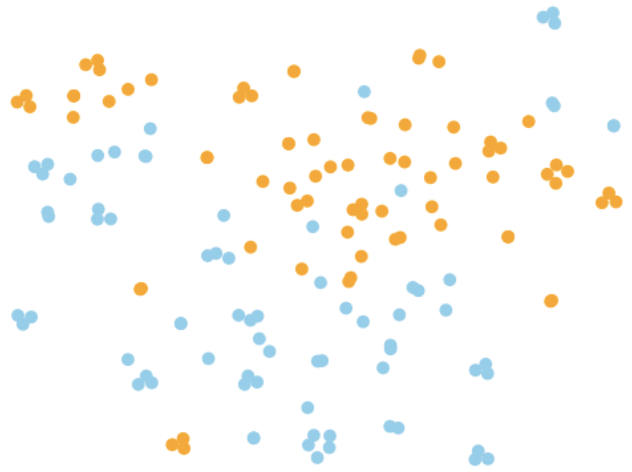}}
  \centerline{\footnotesize{(a) Separation feature from}}
  \centerline{\footnotesize{2048-d ST-SED-SEP}}\medskip
\end{minipage}
\hfill
\begin{minipage}[b]{.48\linewidth}
  \centering
  \centerline{\includegraphics[width=4.0cm]{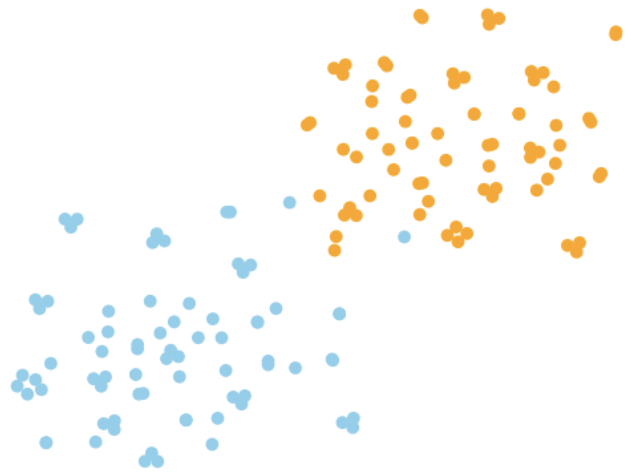}}
  \centerline{\footnotesize{(b) Query embedding from}}
  \centerline{\footnotesize{2048-d ST-SED-SEP}}\medskip
\end{minipage}
\begin{minipage}[b]{0.48\linewidth}
  \centering
  \centerline{\includegraphics[width=4.0cm]{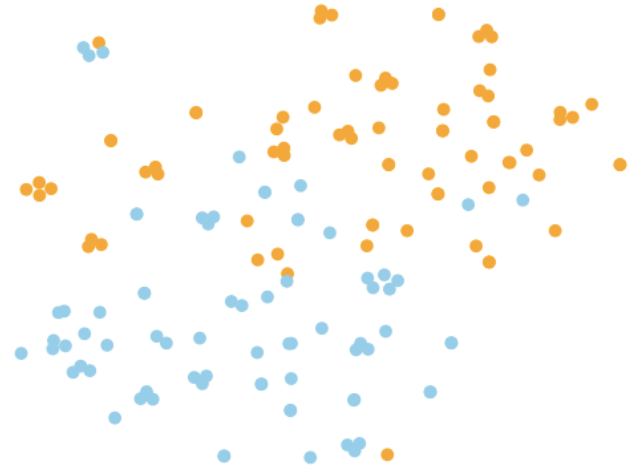}}
  \centerline{\footnotesize{(c) Separation feature from}}
  \centerline{\footnotesize{our 2048-d CaRE-SEP}}\medskip
\end{minipage}
\hfill
\begin{minipage}[b]{0.48\linewidth}
  \centering
  \centerline{\includegraphics[width=4.0cm]{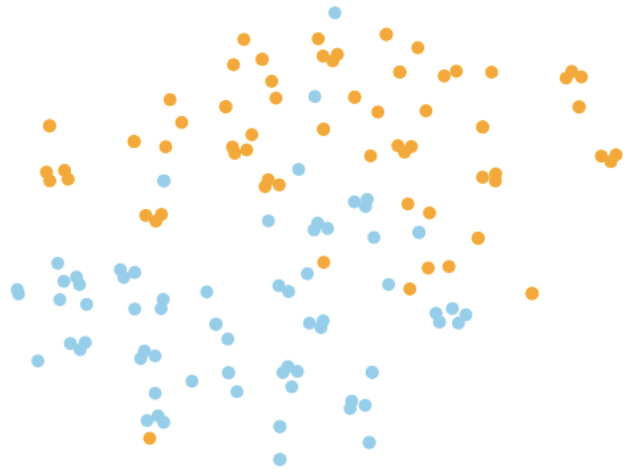}}
  \centerline{\footnotesize{(d) Query embedding from}}
  \centerline{\footnotesize{our 2048-d CaRE-SEP}}\medskip
\end{minipage}
\vspace{-10pt}
\caption{t-SNE visualization of embedding, where different colors represent two categories.}
\label{fig:tsne}
\vspace{-18pt}
\end{figure}

In this section, we aim to analyze the effectiveness of our proposed model from the perspective of feature distribution.
We randomly sample 70 audio samples from each of 2 categories in the AudioSet evaluation set and extract separation feature and query embedding from different models. 
As shown in Fig.\ref{fig:tsne}, we use t-SNE \cite{van2008visualizing} to visualize the distribution of embeddings, with different colors indicating individual categories. 
In the 2048-d ST-SED-SEP (baseline) and our CaRE-SEP, we extract output of the final block in the connector module as the separation feature.


Fig.\ref{fig:tsne}(a) and Fig.\ref{fig:tsne}(c) indicate that separation feature can slightly discriminate categories, with a small inter-class difference. 
However, the query embedding shows more apparent class discriminability in Fig.\ref{fig:tsne}(b), which validates the conjecture that a mismatch exists between the query embedding and separation feature in the baseline.
Fig.\ref{fig:tsne}(d) is almost consistent with Fig.\ref{fig:tsne}(c), which further suggests that our model can generate consistent and relevant query embedding with separation feature.
In addition, we also calculate the KL divergence between the embeddings of two categories. 
The four KL divergences in Fig.\ref{fig:tsne} are (a)4.34, (b)10.37, (c)5.08, (d)5.77, respectively. 
This metric analysis indicates that the distribution between separation feature and query embedding exists mismatches in baseline. While in our CaRE-SEP, their distribution is closer.

Moreover, Fig.\ref{fig:tsne}(b) is obtained from the extra classification model. 
Fig.\ref{fig:tsne}(b) and Fig.\ref{fig:tsne}(a) can indicate the classification features are inconsistent with the separation feature. 
This can also explain the reason for the slight decrease in classification performance when jointly training classification and separation in Section \ref{sec_41}.
\vspace{-8pt}
\subsection{Ablation Study}
\vspace{-5pt}
As demonstrated in Table \ref{res_ablation}, we investigate the impact of query embedding on the performance of separation models from three perspectives. 
$Grad$ refers to the gradient backpropagation during the training (not freezing). 
$Init$ is to initialize the shared encoder with a pre-trained classification model.
$Shared\ encoder$ means that the separation encoder can simultaneously generate query embedding.

We validate the effectiveness of the proposed swin-unet architecture, shared encoder, classification initialization, and gradient backpropagation sequentially. 
First, the proposed swin-unet architecture is tested by comparing models A and G, each of which has an extra classification network \cite{chen2022zero}. 
Model G gets comparable results with the baseline, demonstrating the effectiveness of our proposed model structure.
Second, the shared encoder is tested by comparing models B and C. 
Model B trains the query embedding with gradients in the extra network and performs worse than model C, indicating the importance of encoder sharing.
Third, the initialization of the shared encoder is evaluated by comparing models D and F. 
Model D uses a pre-trained classification model to initialize the shared encoder, which improves the performance by nearly 1 dB compared with model F. This confirms that the classification embedding can help audio source separation.
Fourth, we train the query embedding with gradients in the shared encoder to validate the gradient backpropagation. 
Model E demonstrates a significant increase in mixture SDR when compared to models F and D, indicating that gradient backpropagation is crucial for query embedding.
Moreover, we test model F with a shared encoder to directly generate query embedding, which results in the worst mixture SDR.
This suggests that $Grad$ and $Init$ are indispensable for the shared encoder.

\vspace{-10pt}
\subsection{Results on ESC-50}
\vspace{-22pt}
\begin{table}[htbp]
\centering
\setlength{\tabcolsep}{1.8mm}
\caption{Results on ESC-50}
\begin{tabular}{cc}
\toprule[1pt]
\textbf{Model}
&
\textbf{Mixture-SDR}(dB$\uparrow$)
\\
\hline
2048-d ST-SED-SEP \cite{chen2022zero}
& 9.03
\\
768-d CaRE-SEP
& 9.83
\\
768-d CaRE-CLS+SEP
& 10.34
\\
\bottomrule[1pt]
\end{tabular}
\label{res_on_esc-50}
\vspace{-10pt}
\end{table}

To verify generalization ability, we further evaluate the zero-shot performance on ESC-50. 
Table \ref{res_on_esc-50} shows our CaRE performs better than baseline, they can both demonstrate the effectiveness of consistent and relevant embeddings. 
The performance of joint training is better than CaRE-SEP, which is different from the results on AudioSet. 
This can be explained by that these two datasets are inconsistent. 
The models trained on AudioSet train set perform well in its evaluation set. While directly testing the zero-shot performance on ESC-50, the performance of CLS+SEP is better, indicating labels play an important role in zero-shot generalization. 

%% file: 04_conclusions.tex
\vspace{-8pt}
\section{conclusions}
\vspace{-6pt}
In this study, we propose CaRE-SEP, a consistent and relevant embedding network
for general sound separation.
Different from previous query-based audio separation models which need auxiliary networks to generate query embedding, our CaRE-SEP can make the query embedding and separation feature more consistent and relevant by eliminating structural and informational mismatches in a single model. 
Extensive experiments show the influence of query embedding on separation tasks, as well as significant separation performance achieved by our method. 
Moreover, CaRE-SEP can simultaneously train classification and separation tasks, demonstrating that joint multi-task training is a promising direction in the future.